# Specific Heat vs Field of LiFe$_{1-x}$Cu$_x$As

J. S. Kim[1], G. R. Stewart[1], L.Y. Xing[2], X.C. Wang[2] and C. Q. Jin[2]

[1]Department of Physics, University of Florida, Gainesville, FL 32611-8440
[2]Institute of Physics, Chinese Academy of Sciences, Beijing 100190, China

Abstract: LiFeAs is one of the new class of iron superconductors with a bulk $T_c^{onset}$ in the 15-17 K range. We report on the specific heat characterization of single crystal material prepared from self-flux growth techniques with significantly improved properties, including a much decreased residual gamma, $\gamma_r$, ($\equiv C/T$ as $T\to 0$) in the superconducting state. Thus, in contrast to previous explanations of a finite $\gamma_r$ in LiFeAs being due to *intrinsic* states in the superconducting gap, the present work shows that such a finite residual $\gamma$ in LiFeAs is instead a function of sample quality. Further, since LiFeAs has been characterized as nodeless with multiple superconducting gaps, we report here on its specific heat properties in zero and applied magnetic fields to compare to similar results on nodal iron superconductors. For comparison, we also investigate LiFe$_{0.98}$Cu$_{0.02}$As, which has the reduced $T_c$ of $\approx$ 9 K and $H_{c2}$ of 15 T. Interestingly, although presumably both LiFeAs and LiFe$_{0.98}$Cu$_{0.02}$As are nodeless, they clearly show a non-linear dependence of the electronic density of states ($\propto$ specific heat $\gamma$) at the Fermi energy in the mixed state with applied field similar to the Volovik effect for nodal superconductors. However, rather than nodal behavior, the satisfactory comparison with a recent theory for $\gamma(H)$ for a two isotropic gap superconductor in the presence of impurities argues for nodeless behavior. Thus, in terms of specific heat in magnetic field, LiFeAs can serve as the prototypical multiband, nodeless iron superconductor.

I. Introduction

The iron superconductor LiFeAs, with $T_c$ reportedly varying between 15 and 19 K, was discovered by Wang et al. [1]. Successive characterization by a variety of techniques reveals that LiFeAs can be well described [2] as nodeless. ARPES data reveals four [3] or five [4] Fermi surface pockets, although such a multiplicity of bands is at present beyond the ability of most theories describing physical measurements. A simplified two gap model has been used to fit penetration depth [5] and NMR and NQR [6] measurements.

Since the consensus is that LiFeAs is nodeless, it was of interest to examine the field dependence of the specific-heat-determined electronic density of states $\propto \gamma$, where $\gamma$ is the coefficient of the linear-with-temperature term in the low temperature specific heat, $C=\gamma T + \beta T^3$. In nodal superconductors in the superconducting state, $\gamma(H) \propto H^{1/2}$ at low fields, followed by $\gamma \propto H^1$ (Volovik behavior) as has been recently seen [7] in $BaFe_2(As_{0.7}P_{0.3})_2$ and overdoped $Ba(Fe_{1-x}Co_x)_2As_2$ [8[. Thus, in nodeless LiFeAs, we would <u>a priori</u> expect qualitatively different $\gamma(H)$ behavior.

A further reason to investigate LiFeAs concerns the correlation between the discontinuity in the specific heat at $T_c$, $\Delta C$, and $T_c$ first discovered by Bud'ko, Ni, and Canfield [9] (hereafter 'BNC'), where they found $\Delta C/T_c \propto T_c^2$ for 14 different compositions of 122 structure iron based superconductors. LiFeAs samples that have been characterized to date, despite having similar $T_c^{midpoint}$ determined from bulk specific heat, of 14.7 to 16 K, had large variations in both $\Delta C/T_c$ and residual $\gamma_r$ ($\equiv C/T$ as $T\rightarrow 0$). Specifically, $\Delta C/T_c$ ranges from 7.65 mJ/molK$^2$ [10], to 12.4 mJ/molK$^2$ [11], to $\approx$20 mJ/molK$^2$ in ref. 12 while residual $\gamma_r$ values (instead of $\gamma_r \approx 0$ in a 100 % gapped superconducting sample) range from 7.7 to 0.9 to 18.7 mJ/molK$^2$ respectively in the three references. (Note that specific heat data were only taken down to 2 K in ref. 11 resulting in some inaccuracy to their $\gamma_r$ value.)

The samples made from self flux characterized in the present work have almost vanishing $\gamma_r$. Thus it was considered valuable to determine $\Delta C/T_c$ for these higher quality LiFeAs crystals. $\Delta C/T_c$ for LiFe$_{0.98}$Cu$_{0.02}$As, $T_c^{mid}$=9.3 K, was also measured for comparison.

II. Experimental

Single crystals of LiFeAs and LiFe$_{0.98}$Cu$_{0.02}$As were prepared by self flux methods. FeAs and Fe$_{0.98}$Cu$_{0.02}$As were pre-synthesized by reacting the mixture of elemental powder at 750°C for 20 hours in an evacuated quartz tube; Li$_3$As by reacting Li lumps with As powder at 600°C for 10 hours. The Li$_3$As, FeAs (or Fe$_{0.98}$Cu$_{0.02}$As), and As powders were mixed according to the element ratio of Li(Fe$_{1-x}$Cu$_x$)$_{0.3}$As (x=0, 0.02). The powder mixture was then pressed into a pellet in an alumina oxide tube and sealed in a Nb tube under 1 atm of Argon gas and then sealed in an evacuated quartz tube. The sealed quartz tube was heated up to 1100°C for 10 h and then cooled down to 700°C at a rate of 5°C per hour. Crystals with a size up to 10 mm × 6 mm × 0.5 mm were obtained. The whole preparation work was carried out in a glove box protected with high purity Ar gas.

A study was done as to what attachment method would react least with the known-to-be-reactive LiFeAs crystals. Both GE7031 and Wakefield grease were found to give minimal degration in $T_c^{onset}$ and transition width if applied within 1 day or less of cooling down to low temperatures. After a slight degradation after one day (comparable to the result of ref. 12 which used GE7031), further storage at room temperature showed stable $T_c^{onset}$ and $\Delta T_c$.

Specific heat in zero and applied magnetic fields up to 12 T was measured down to 0.4 K according to established techniques [13]. The sample platform was reinforced to increase the strength of the support wires by the addition of two crossed hollow cylinders of kapton, 0.0065" OD, 0.00075" wall, attached to the platform with epoxy [14]. Due to the existence of an unknown magnetic impurity phase the sample experienced a large magnetic force in magnetic fields >9 T. Such technical difficulties in determining $\gamma(H)$ have been seen in other systems, e. g. Co-doped $BaFe_2As_2$ [8]. This force caused the sample to fly off the Wakefield grease bonding to the sample platform. The sample was reattached with the more secure GE7031 varnish, and then further measured up to 12 T. At some larger field, the force would exceed the support platform's failure strength and thus data collection was halted after 12 T.

III. Results and Discussion

Fig. 1 shows the $\Delta C/T_c$ determinations for both LiFeAs and $LiFe_{0.98}Cu_{0.02}As$ and Fig. 2 shows the low temperature specific heat down to 0.4 K to determine $\gamma_r$. The previous values for $\Delta C/T_c$ for LiFeAs were 7.65 mJ/molK$^2$, ($T_c^{midpoint}$ =15 K from the specific heat, transition width $\Delta T_c \cong 3$ K, $\gamma_r \cong 7.7$ mJ/molK$^2$) (ref. 10); 12.4 mJ/moleK$^2$ ($T_c$=14.7 K, $\Delta T_c \cong 1.3$ K, $\gamma_r \cong 0.9$ mJ/molK$^2$) (ref. 11) and $\approx$20 mJ/moleK$^2$ ($T_c$=16.8 K, $\Delta T_c \cong 2.5$ K, $\gamma_r$=18.7 mJ/molK$^2$) (ref. 12). These results compare to the present work's (see Fig. 1) $\Delta C/T_c$ =16.7 mJ/molK$^2$, with a transition width $\Delta T_c$ of $\approx$1 K while Fig. 2 shows that $\gamma_r$=0.4 mJ/molK$^2$.

Thus, we can draw two conclusions from these results. First, as clear from the data in Figs. 1 and 2 (considering the narrowness of the transition and the low, 0.4 mJ/molK$^2$ $\gamma_r$ for the parent LiFeAs compound), the correct $\Delta C/T_c$ value for pure LiFeAs is close to 17 mJ/molK$^2$. $\Delta C/T_c$ for the lower transition temperature, somewhat broader $\Delta T_c$ Cu-doped sample is 5 mJ/molK$^2$. These values agree well with the overall $\Delta C/T_c \propto T_c^2$ trend for the iron based superconductors established [9] by Bud'ko, Ni and Canfield and later confirmed [15] by Kim et al. for a wider spectrum of samples.

A second conclusion that can be drawn from Fig. 2 is that a large $\gamma_r$ such as reported by refs. 10 and 12 is *not* intrinsic to LiFeAs. Although there is a low temperature Schottky anomaly upturn in C/T in Fig. 2 (as has been seen in other iron based superconductors, see e. g. refs. 16-17), this does not affect the accuracy of the extrapolation of $\gamma_r$ to only 0.4 mJ/molK$^2$. Thus, we suggest that in LiFeAs in particular, and in other iron based superconductors in general, that the size of the residual gamma is proportional to the sample quality. This is consistent with the evidence in refs. 17 and 18 that $\gamma_r$ decreases with annealing in Ba(Fe$_{1-x}$Co$_x$)$_2$As$_2$. Thus, the correct explanation of this residual gamma is that it is more likely due to a non-superconducting fraction rather than states in the gap due to strong unitary scattering as has been theorized by some works to explain the presence of substantial $\gamma_r$ values.

Now let us consider the field dependences of C/T of LiFeAs and LiFe$_{0.98}$Cu$_{0.02}$As, where the specific heat data vs temperature as a function of field are plotted in Figs. 3 and 4 respectively. As has been discussed thoroughly in ref. 8, we present two methods to track $\gamma$ (where C/T=$\gamma$ as T→0) as a function of field. The first method is a fit of the C/T data to C/T=$\gamma$ + ß$T^2$ + δ$T^4$ in a temperature range above any anomalies or upturns (the higher field, low temperature data for LiFe$_{0.98}$Cu$_{0.02}$As in Fig. 4 show an upturn due to field splitting of the nuclear hyperfine levels). The second method is a fit of the C/T data in the close neighborhood of 2 K, since C/T(2 K) is proportional to γ. This latter proportionality is readily apparent when both methods of tracking γ vs H are shown in Figs. 5, 7-8.

Consider first γ and C/T(2 K) vs H for LiFeAs, shown in Fig. 5. γ can be fit to a simple power law of field, giving a field dependence of γ∝H$^{0.66}$ over the entire field range of measurement. Although these γ vs H data may look 'Volovik-like' (i. e. indicative of nodes or deep minima), in fact – as shown in Fig. 6 – a two isotropic band fit like that proposed by Y. Bang (which in ref. 19 only extends up to H/H$_{c2}$=0.4 but see ref. 10) gives a good fit of the data.

Similarly, γ and C/T(2 K) vs H for LiFe$_{0.98}$Cu$_{0.02}$As, shown in Fig. 7, also follow a simple power law, where γ∝H$^{0.57}$. Since there was no parasitic ferromagnetic phase to cause the sample to fly off the platform, these data were measured up past H$_{c2}$≈15 T up to 22 T. In the high field normal state data it is apparent that the normal state γ is slightly field dependent. Fig. 8 shows this normal state γ field dependence factored into the fit of the γ vs H data in the superconducting state to give a slightly altered (0.54 vs 0.57) field exponent. This power law is reminiscent of Volovik behavior in YBCO, where ref. 20 found that γ varied as H$^{1/2}$ indicative of nodes in the superconducting gap function. However, just as for LiFeAs, Fig. 6, a fit (not shown) of these γ vs H data for LiFe$_{0.98}$Cu$_{0.02}$As to the 2 isotropic band fit of Y. Bang [19] gives a convincingly good agreement between the fit and the data.

Thus, these specific heat data in field in a superconductor system known to be nodeless serve as a warning that $\gamma \propto H^\alpha$, $\alpha$ near 0.5-0.7, is *not* necessarily an indication of nodal or deep minima behavior in the superconducting gap. This sub-linear behavior of $\gamma$ with H has also been reported [21] in the 122 defect structure $Rb_{1-x}Fe_{2-y}Se_2$, $T_c$=32 K, which is also believed to be nodeless.

IV. Conclusions

The specific heat in zero and applied field of single crystals of LiFeAs and $LiFe_{0.98}Cu_{0.02}As$ showed good agreement between $\Delta C/T_c$ and the published trend [9,15] vs $T_c$ for all iron based superconductors. The measured residual $\gamma$, $\gamma_r$, for pure LiFeAs is only 0.4 mJ/molK$^2$, which suggests the conclusion that previously observed larger values are not characteristic of the intrinsic properties. Thus, as a speculation, such finite residual gamma values in other iron based superconductors, e. g. in Co-doped $BaFe_2As_2$, may also be merely due to sample quality issues. The field dependence of the specific heat $\gamma$ of both LiFeAs and $LiFe_{0.98}Cu_{0.02}As$ obeys approximately an $H^{0.6\pm0.05}$ power law. Although such a field dependence is reminiscent of the Volovik effect and its inference of nodal superconductivity, in these known-to-be-nodeless materials this $\gamma(H)$ behavior is instead indicative of the behavior of two (or more) approximately isotropic bands as proposed by Y. Bang [19].

Acknowledgements: The authors wish to thank Jim Obrien and Randy Black of Quantum Design for assistance with the kapton tube strengthening of the sample platform.

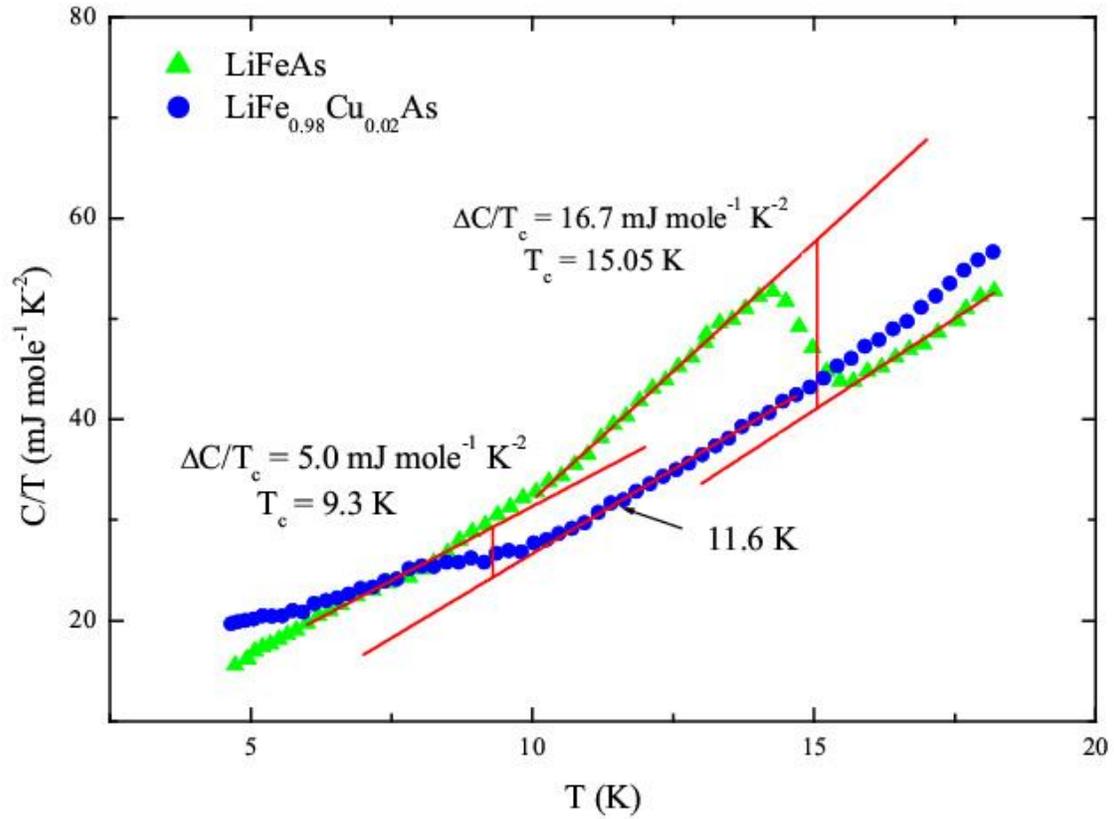

Fig. 1 (color online) Specific heat divided by temperature, C/T, vs temperature for single crystal LiFeAs and LiFe$_{0.98}$Cu$_{0.02}$As expanded around T$_c$ to show the discontinuity at the superconducting transition, ΔC. The equal area constructions shown in red are discussed in ref. 2 and represent idealized sharp transitions. The bulk transition width, ΔT$_c$, for the undoped LiFeAs is as narrow or narrower than previously seen.

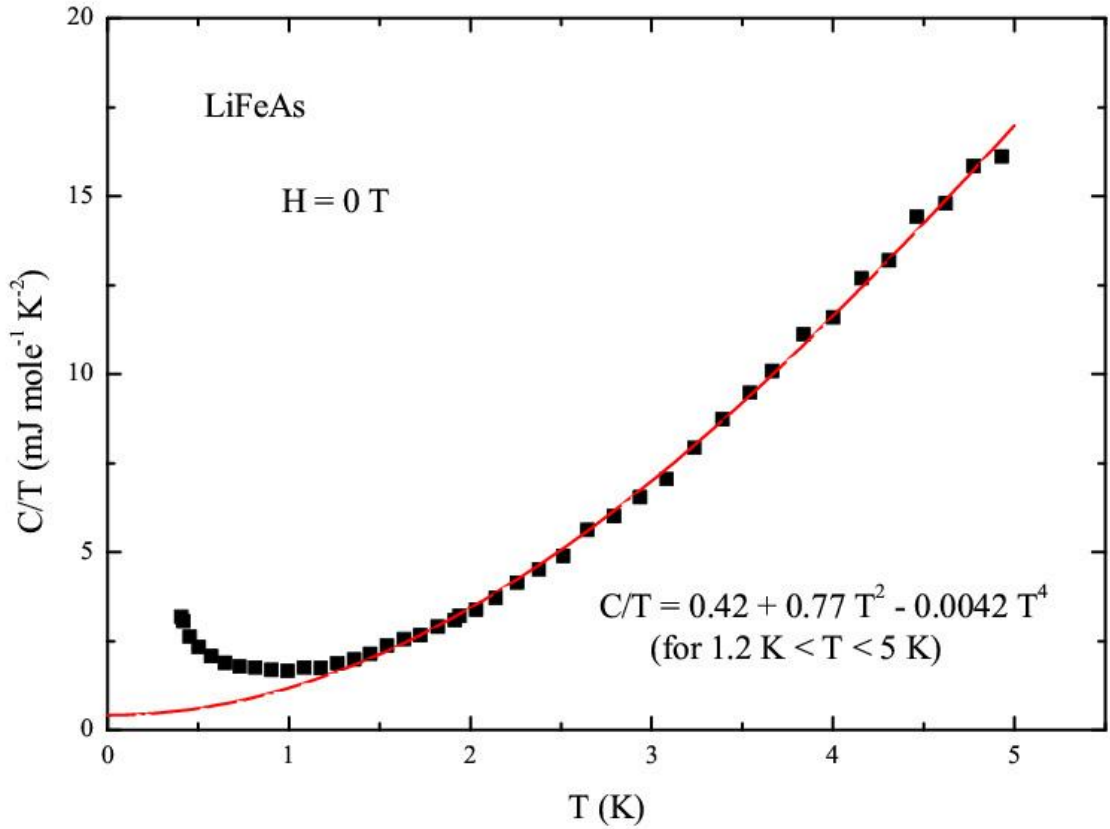

Fig. 2 (color online) Low temperature C/T vs temperature for single crystal LiFeAs. The low temperature upturn below 1.3 K is an unknown minority phase Schottky anomaly as seen in other iron based superconductors [16-17]. Note the extremely low extrapolated residual gamma value, $\gamma_r$, of 0.4 mJ/molK$^2$.

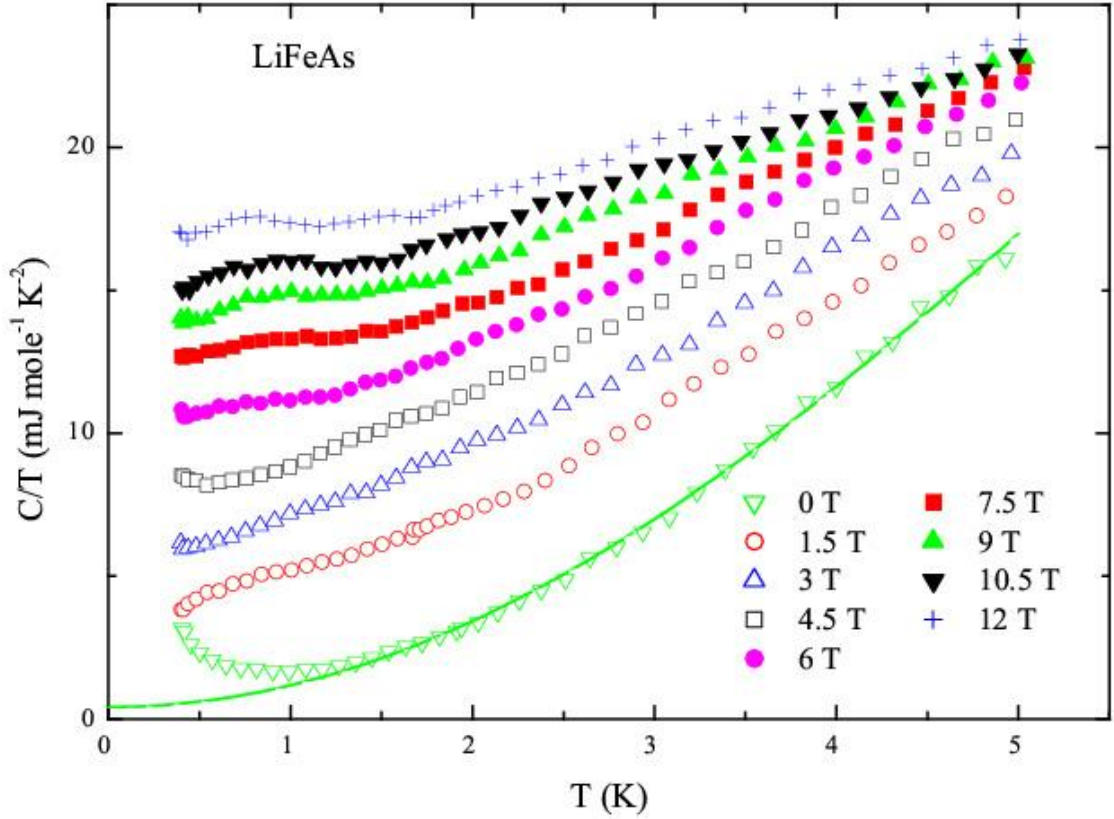

Fig. 3 (color online) C/T versus temperature in fields up to 12 T of single crystal LiFeAs with field aligned in the c-axis direction. γ values (≡ C/T as T→0) are determined by fitting the data from 2 to 5 K, avoiding the small anomaly in C/T around 1 K that grows with increasing field.   For comparison to γ(H), smoothed values of C/T (2 K, H) are also determined by fitting the data in the vicinity of 2 K as discussed in the text and refs. 7-8.

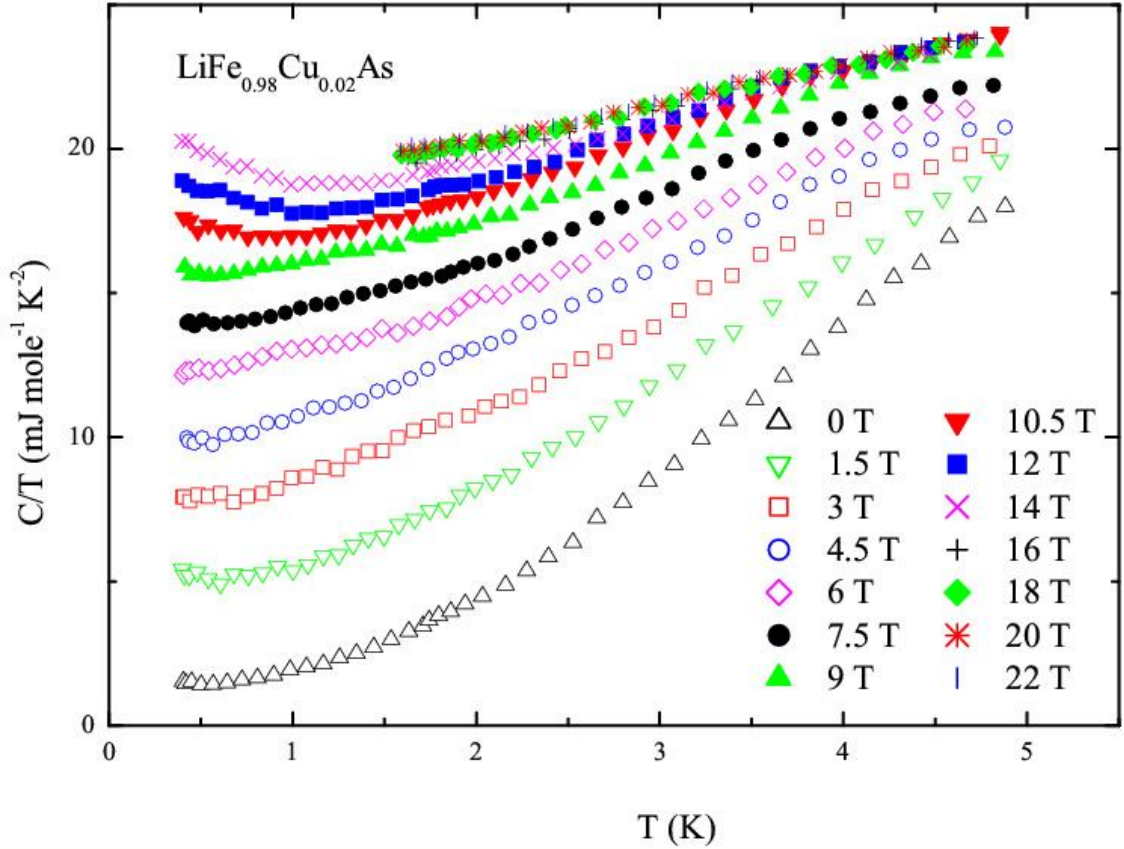

Fig. 4 (color online) C/T versus temperature in fields up to 12 T of LiFe$_{0.98}$Cu$_{0.02}$As, field || c-axis. γ values (≡ C/T as T→0) are determined by fitting the data from 2 to 5 K to C/T=γ + ß$T^2$ + δ$T^4$, avoiding the upturn at low temperatures due to field splitting of the nuclear levels that grows with increasing field.   Note the larger upturn for the Cu-doped sample in 12 T compared to pure LiFeAs in Fig. 3.  This is not just due to the extra contribution to C/T(H) from the 2% Cu, but is presumably (see an exhaustive discussion in ref. 8) primarily due to the lower electronic density of states at the Fermi energy, N(0), in LiFeAs at 12 T.  This N(0) is responsible for coupling the nuclear levels to the rest of the lattice, and is larger at 12 T in the smaller H$_{c2}$ LiFe$_{0.98}$Cu$_{0.02}$As. For comparison to γ(H), smoothed values of C/T (2 K, H) are also determined by fitting the data in the vicinity of 2 K as discussed in the text and refs. 7-8.

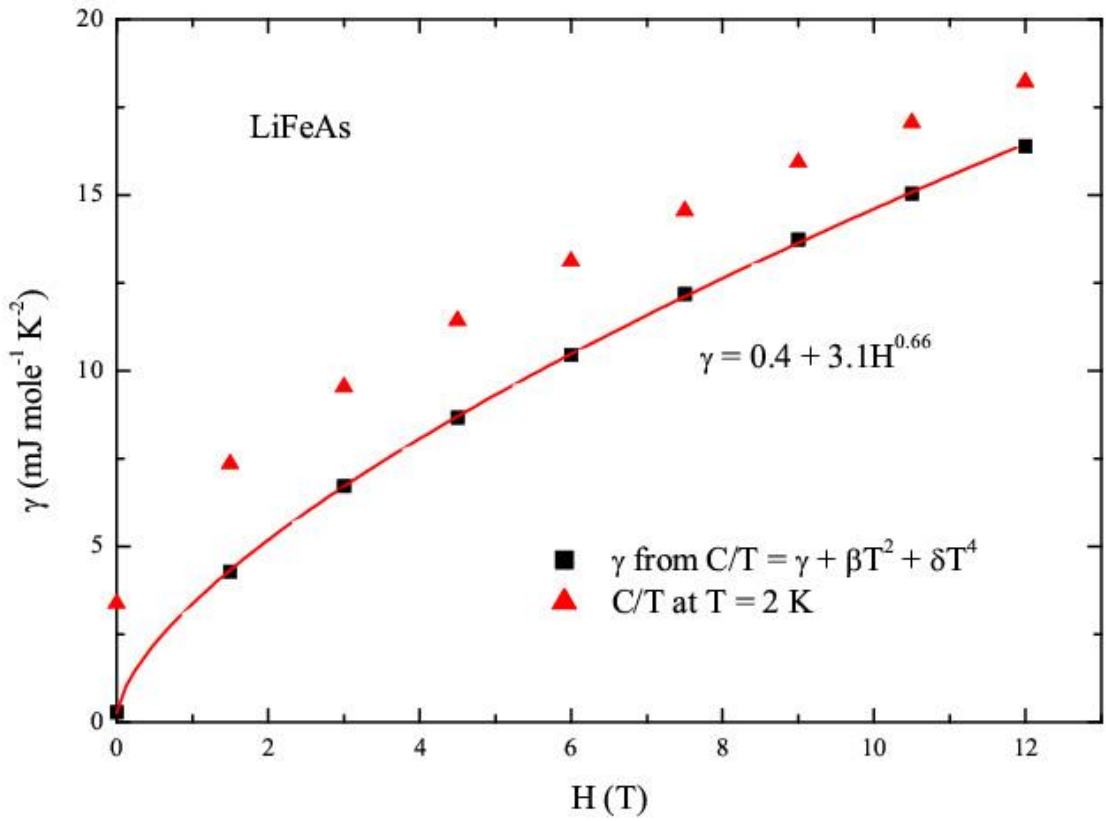

Fig. 5 (color online) Specific heat γ from a three term polynomial fit to the 2-5 K data in Fig. 3 vs H for LiFeAs, as well as C/T (2 K) vs H for comparison. The two metrics for the field dependence of γ agree very well, and result in γ∝$H^{0.66}$.

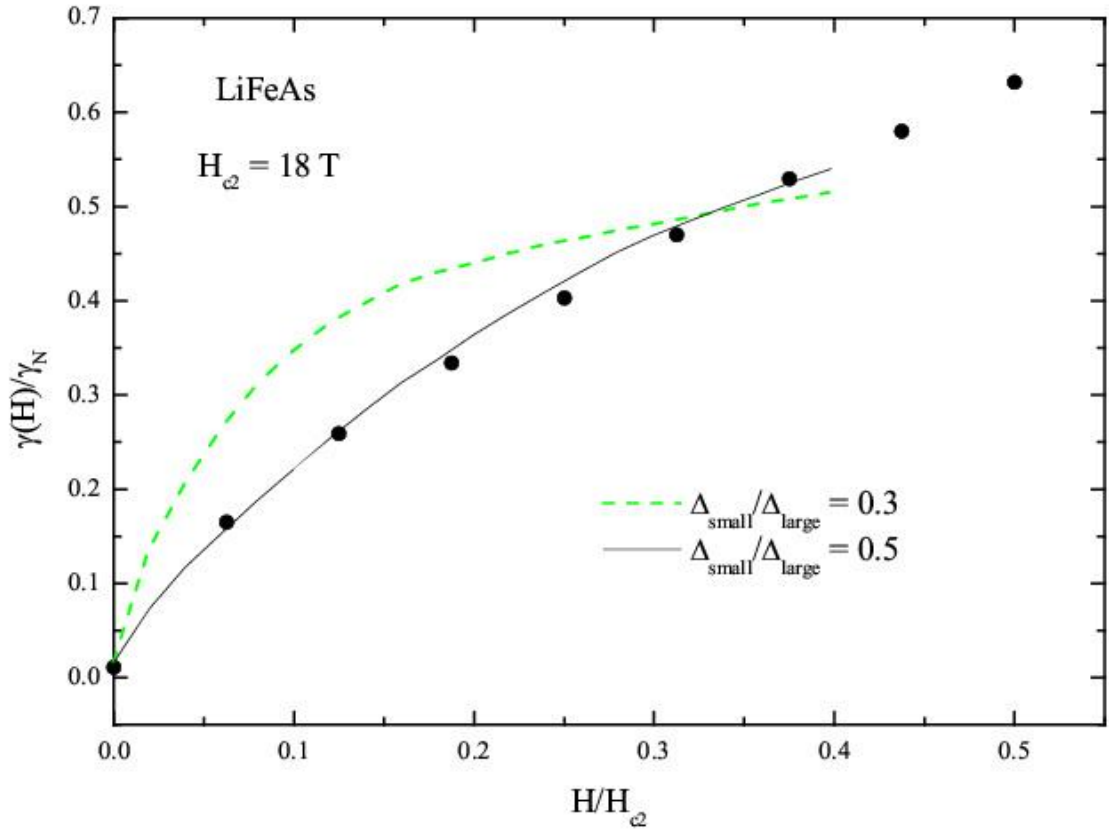

Fig. 6 (color online)  Shown here is a fit (obtained from ref. 19) to the γ(H) data from Fig. 5 for LiFeAs based on a two isotropic band model, where the theory in ref. 19 is only up to $H/H_{c2} = 0.4$.  Although ARPES data find more than 2 bands in the iron based superconductors, the ratio of the band gaps of 0.5, $\Delta_{small}/\Delta_{large}$, is not inconsistent with the ARPES-determined band gaps [3-4].

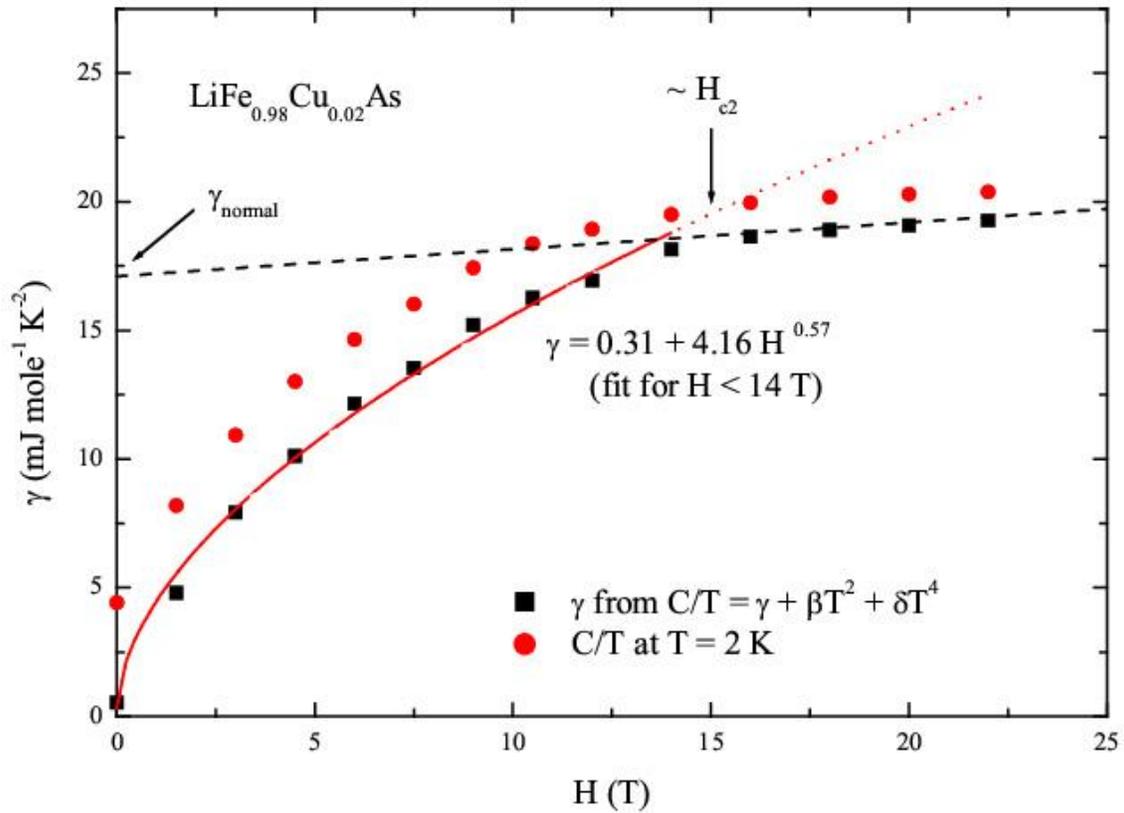

Fig. 7 (color online) Specific heat γ from a three term polynomial fit to the 2-5 K data in Fig. 4 vs H for LiFe$_{0.98}$Cu$_{0.02}$As, as well as C/T (2 K) vs H for comparison. These field data are up to 22 T, well above H$_{c2}$≈15 T. The two metrics for the field dependence of γ, just as for LiFeAs in Fig. 5, agree very well, and result in γ∝H$^{0.57}$. The fact that the normal state γ appears to be a slight function of field implies that the power law dependence of γ in the superconducting state could be altered by the field dependence of the specific heat of the normal state cores. This issue is addressed below in Fig. 8.

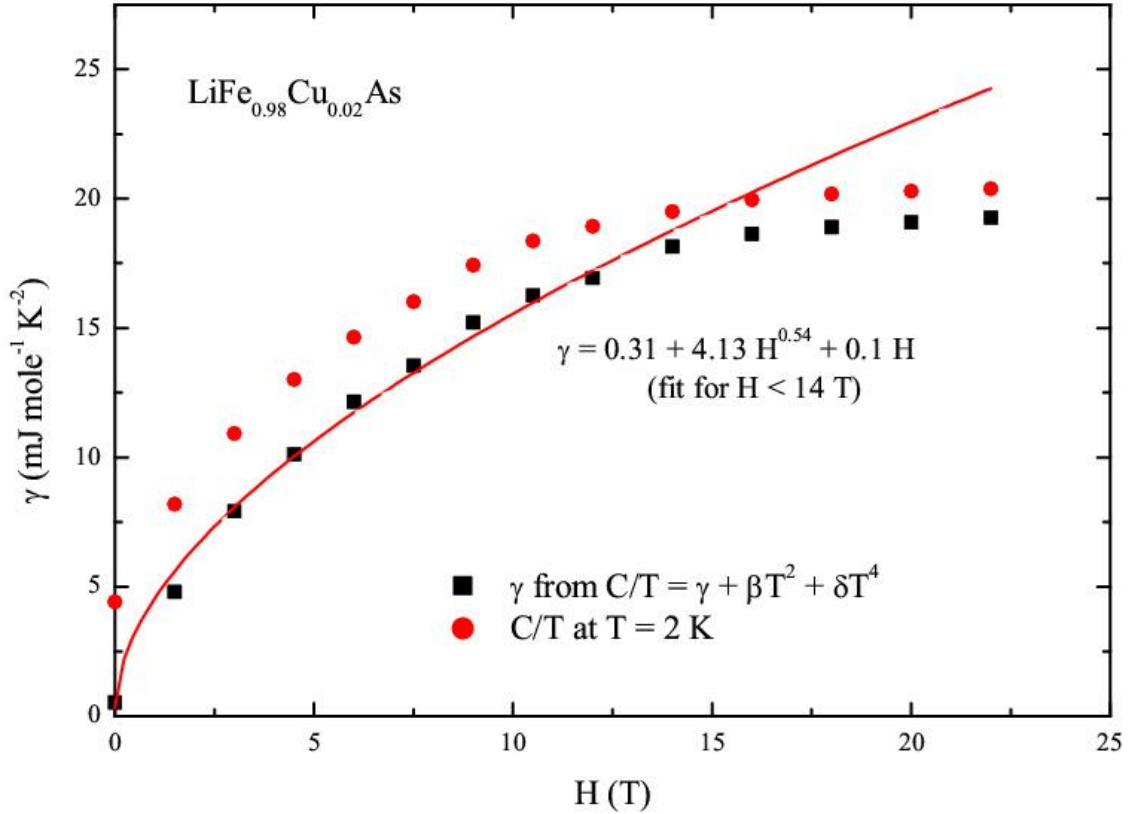

Fig. 8 (color online) γ and C/T (2 K) vs field in LiFe$_{0.98}$Cu$_{0.02}$As, with the power law fit adjusted for the field dependence of the normal state γ determined in Fig. 7. As may be seen, the change in the power law exponent is within the error bar, from 0.57 without adjustment to 0.54 with.